\documentclass[aps,prl,twocolumn,superscriptaddress,amsmath,amssymb]{revtex4}
\usepackage{graphicx}
\usepackage{amsmath}

\begin{document}

\title{Universal Trimers induced by Spin-Orbit Coupling in Ultracold Fermi Gases}
\author{Zhe-Yu Shi}
\email{shizy07@mails.tsinghua.edu.cn}
\affiliation{Institute for Advanced Study, Tsinghua University, Beijing, 100084, China}
\author{Xiaoling Cui}
\email{xlcui@iphy.ac.cn}
\affiliation{Institute for Advanced Study, Tsinghua University, Beijing, 100084, China}
\affiliation{Beijing National Laboratory for Condensed Matter Physics, Institute of Physics, Chinese Academy of Sciences, Beijing, 100190, China}
\author{Hui Zhai}
\email{hzhai@tsinghua.edu.cn}
\affiliation{Institute for Advanced Study, Tsinghua University, Beijing, 100084, China}

\date{\today}
\begin{abstract}

In this letter we address the issue how synthetic spin-orbit (SO) coupling can strongly affect three-body physics in ultracold atomic gases. We consider a system which consists of three fermionic atoms, including two spinless heavy atoms and one spin-$1/2$ light atom subjected to an isotropic SO coupling. We find that SO coupling can induce universal three-body bound states with negative s-wave scattering length at a smaller mass ratio, where no trimer bound state can exist if in the absence of SO coupling. The energies of these trimers are independent of high-energy cutoff, and therefore they are universal ones. Moreover, the resulting atom-dimer resonance can be effectively controlled by SO coupling strength. Our results can be applied to systems like ${}^6$Li and ${}^{40}$K mixture.

\end{abstract}
\maketitle

``Universal phenomenon" refers to observations independent of short-range or high energy details, which is one of the most beautiful and charming parts of physics. Universal physics not only emerges in interacting many-body systems but also exists in quantum mechanical few-body problems. Cold atoms system, because of its diluteness, is an ideal platform to investigate various intriguing phenomena of few-body systems. For instance, Efimov trimer with universal scaling factor \cite{Efimov,Brateen} has been extensively studied experimentally \cite{Efimov_Exp1,Efimov_Exp2,Efimov_Exp3,Efimov_Exp4,Efimov_Exp5,Efimov_Exp6,Efimov_Exp7,Efimov_Exp8}. Another type of trimer whose energy is universal has also been predicted by Kartavtsev and Malykh \cite{KM}.

On the other hand, thanks to fast experimental developments \cite{Spielman_exp1,Spielman_exp2,Shuai,Spielman_exp3,Jing,MIT,Chuanwei,Spielman_exp4,Shuai_2013,Spielman_2013,Jing_2013}, synthetic spin-orbit (SO) coupling recently emerges as one of the most exciting research directions in cold atom physics \cite{review}. Among many profound effects of SO coupling, one distinct factor is that certain types of SO coupling can dramatically change the two-body physics. For instance, with Rashba-type SO coupling, because the low-energy density-of-state is enhanced to a finite constant, any small attractive interaction between atoms can support a two-body bound state in three-dimension, and the binding energy increases with the strength of SO coupling \cite{Vijay}.
Consequently, this two-body result dramatically changes many-body physics in the scenario of BEC-BCS crossover for spin-$1/2$ fermions \cite{Yu_Zhai,Vijay_MF,Hu}, where the superfluidity is greatly enhanced by SO coupling even in the far BCS side \cite{Yu_Zhai}.

The dramatic effect of SO coupling in two-body problem and its profound consequence naturally raises the question whether similar significant manifestation also exists in a three-body problem. However, so far three-body problems with SO coupling have not been studied in cold atom content, though historically there were some related studies in investigating nucleus \cite{nucleus1,nucleus2,nucleus3}.  In this work we study a three-fermion problem which consists of two heavy fermionic $\alpha$-atoms with mass $M$ and one light fermionic $\beta$-atom with mass $m$, and $\alpha$- and $\beta$-atom interact via a zero-range $s$-wave interaction in the vicinity of two-body scattering resonances. $\alpha$-atom is spinless and $\beta$-atom is spin-$1/2$. As the first attempt to demonstrate rich physics of SO coupling in the few-body cold atoms system, we consider a simply case that only $\beta$-atom is subjected to an isotropic SO coupling \cite{spielman_3d_1,spielman_3d_2}. This is realistic for cold atoms system, since synthetic SO coupling for atoms is induced by atom-light (or atom-magnetic field) interaction which can be selectively applied to certain species. For instance, we can consider a mixture of two-component ${}^6$Li with single component ${}^{40}$K, and the (pseudo)-spin of ${}^6$Li is coupled to its momentum \cite{magnetic}.

Indeed, we find that SO coupling leads to intriguing new physics in this three-body system. The most significant finding is that when $M/m \gtrsim 5.92 $ (satisfied by ${}^6$Li and ${}^{40}$K mixture), SO coupling can induce universal trimer state whose energy is independent of short-range parameter. Such trimers can exist at the  negative scattering length side --- a regime where universal trimer can never exist in the absence of SO coupling. Moreover, the locations of three-body resonances are tunable by the strength of SO coupling. This result reveals a unique manifestation of SO coupling in dilute quantum gases and also adds new control knob to three-body system. Potentially it can also shed light on few-body system of nucleus where SO coupling is inevitable.

Before proceeding, we shall first briefly review the known results for such an $\alpha-\alpha-\beta$ system without SO coupling. Two types of trimer states have been found before. First, when $M/m>13.6$, Efimov trimer emerges in both sides nearby resonance. The energy of Efimov trimer is not universal since it depends on the high-energy cutoff known as three-body parameter, while the energies of two successive trimers obey a universal scaling behavior \cite{Efimov}. Secondly, when $8.17<M/m<13.6$, there exists another type of trimer named as ``Kartavtsev-Malykh" trimer, whose energy is universal (i.e. independent of any high-energy cutoff) \cite{KM}. Since the $s$-wave scattering length $a$ is the only length scale, the trimer energy has to simply scale with two-body binding energy. Thus, such universal trimer appears only for positive $a$ when a two-body bound state exists. Due to anti-symmetrization of two $\alpha$-atoms, both two types of trimer states have total angular momentum $L=1$.

\textit{Model.} Our system is described by Hamiltonian $\hat{H}=\hat{H}_0+\hat{U}$,
\begin{align}
\hat{H}_0&=\frac{\mathbf{p_1}^2}{2M}+\frac{\mathbf{p_2}^2}{2M}+\frac{(\mathbf{p_3}-\lambda\hat\sigma)^2}{2m}
 \\
\hat{U}&=[g\delta(\mathbf{r_1-r_3})+g\delta(\mathbf{r_2-r_3})]\mathbf{I},
\end{align}
in which ${\bf p}_{1,2}$(${\bf r}_{1,2}$) refers to the momentum (position) of two $\alpha$-atoms, and ${\bf p}_{3}$(${\bf r}_{3}$) is for $\beta$-atom. $\hat{\sigma}$ is the spin of $\beta$-atom, which couples to its momentum via a three-dimensional isotropic SO coupling $\lambda{\bf p}\cdot\hat{\sigma}$ where ${\bf p}=(p_x,p_y,p_z)$ and $\hat{\sigma}=(\sigma_x,\sigma_y,\sigma_z)$. Without loss of generality, we take $\lambda>0$. Proposals for realizing such a SO coupling have been presented in Ref. \cite{spielman_3d_1,spielman_3d_2}. $s$-wave contact interaction $\hat{U}$ only takes place between $\beta$-atom and $\alpha$-atom, and the interaction strength is assumed to be independent of the spin-index of $\beta$-atom, where ${\bf I}$ in $\hat{U}$ denotes identity operator acting on the spin space of $\beta$-atom.  $g$ is related to $a$ by $\frac{1}{g}=\frac{Mm}{2\pi(M+m) a}-\frac{1}{\Omega}\sum_{\mathbf{k}}\frac{2Mm}{(M+m)k^2}$, where $\Omega$ is the volume. It has been shown that this relation will not be changed by SO coupling, as long as $1/\lambda$ is much larger than the range of inter-atomic potential \cite{Cui, Peng2,Yu}.

To address the three-body bound state, we should first solve the two-body problem with one $\alpha$- and one $\beta$- atom to determine the atom-dimer threshold, which can be carried out quite straightforwardly with Lippman-Schwinger equation \cite{supple}. Although our case differs from previous studies of two-body problem with SO coupling \cite{Vijay,spielman_3d_1,spielman_3d_2,Cui,Peng2,Yu,Peng,Peng3,Vajay2,You} where both two atoms are subjected to SO coupling, the results are quite similar to previous cases with Rashba or three-dimensional isotropic SO coupling, i.e. for any mass ratio $M/m$ and for all $a$, a two-body bound state with zero center-of-mass momentum exists, and the dimer energy $E_\text{d}<0$ \cite{Vijay,Yu_Zhai,spielman_3d_1}. The physical reason is also attributed to the enhancement of density-of-state of $\beta$-atom, which diverges at zero-energy.

For the same three-body system without SO coupling, the total orbital angular momentum $\mathbf{L}$ is a good quantum number and most previous calculations focus on the lowest bound states in $L=1$ channel. After introducing spin degrees of freedom for $\beta$-atom, these bound states are always six-fold degenerate. In the presence of SO coupling, these states would split into two channels with different total angular momentum $\mathbf{J=L+S}$. They are two states with $J=1/2$ and four states with $J=3/2$.

\begin{figure}[t]
\includegraphics[width=3.1 in]
{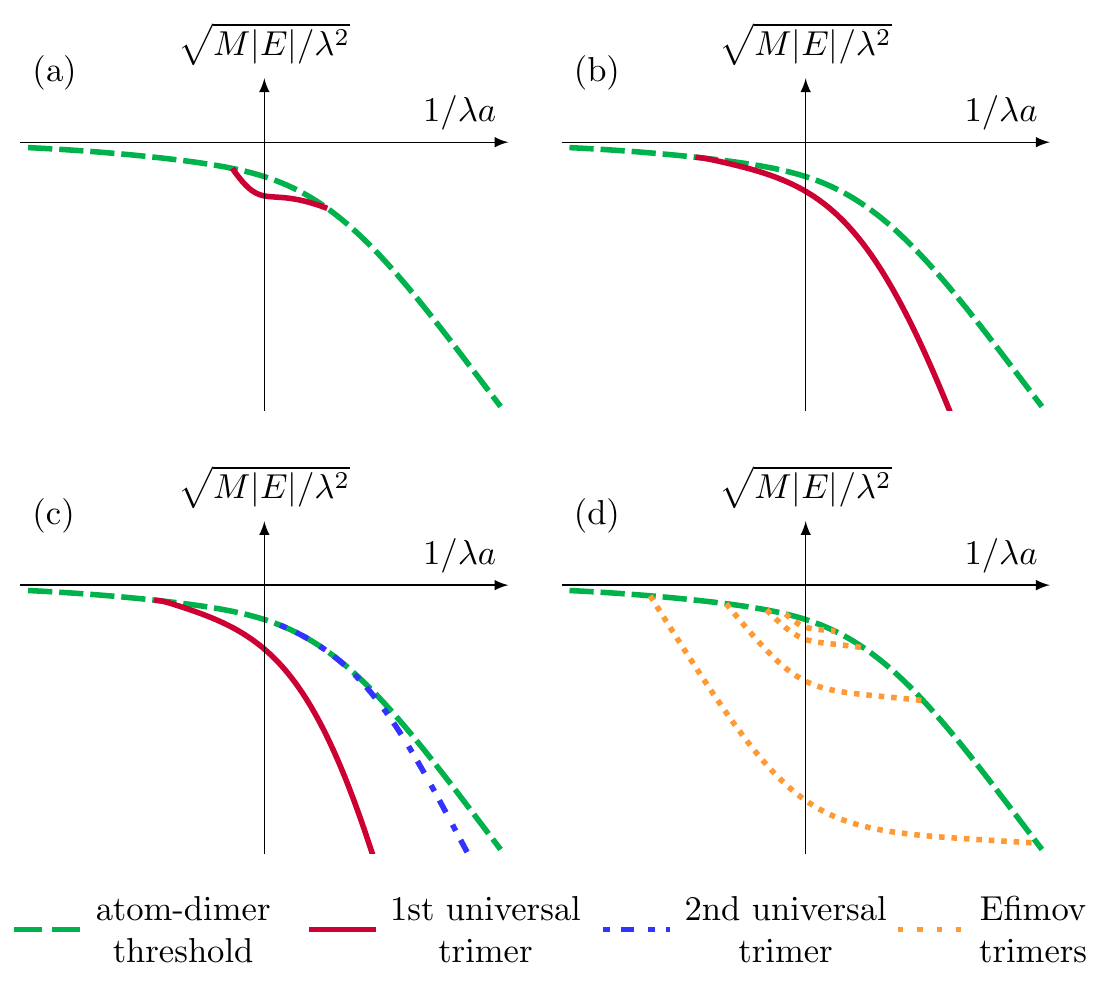}
\caption{Schematic of atom-dimer threshold (green dashed line) and trimer energy in presence of SO coupling for $6.5<M/m<8.17$(a), $8.17<M/m<12.9$(b), $12.9<M/m<13.6$(c) and $13.6<M/m$(d). Red solid line in (a-c) represents the universal trimer with lowest energy. Blue dash-dotted line in (c) represents the second universal trimer; and yellow dotted lines in (d) represent Efimov trimers. This is a schematic plot in order to highlight main features. The actual numbers are shown in Fig. 2. \label{schematic}}
\end{figure}

\textit{Solving Three-body Problem.} Generally, we assume the three-body wave function as
\begin{equation}
|\Psi\rangle=\sum_{{\bf p},{\bf q},\sigma}\Psi_{\sigma}({\bf q},\mathbf{ K_0- p},{\bf p-q})\hat{\alpha}^\dag_{{\bf q}}\hat{\alpha}^\dag_{\mathbf {K_0 -p}}\hat{\beta}^\dag_{\sigma,{\bf p-q}}|0\rangle,
\end{equation}
where $\hat{\alpha}^\dag$ and $\hat{\beta}^\dag$ are creation operators for $\alpha$-atom and $\beta$-atom, respectively, and $\sigma=\uparrow,\downarrow$ is the spin index of $\beta$-atom. Introducing an auxiliary function
$f_\sigma(\mathbf{p})=g\sum_{\mathbf{q}}\Psi_\sigma(\mathbf{q,K_0-p,p-q})$, we can reach following integral equation for $f_\sigma({\bf q})$:
\begin{eqnarray}
f_\sigma(\mathbf{k})=g\sum_{\mathbf{p},\sigma^\prime}G_{\sigma\sigma^\prime}(E;\mathbf{p,K_0-k,k-p}) \nonumber\\
\times[f_{\sigma^\prime}(\mathbf{k})- f_{\sigma^\prime}(\mathbf{K_0-p})],\label{int_eqn1}
\end{eqnarray}
where
\begin{eqnarray}
G_{\sigma\sigma^\prime}(E;\mathbf{k_1,k_2,k_3})=\langle\mathbf{k_1,k_2,k_3};\sigma|\frac{1}{E-H_0}|\mathbf{k_1,k_2,k_3};\sigma^\prime\rangle \nonumber
\end{eqnarray}
is the Green's function in momentum space \cite{tan,petrov2,cui_tmatrix}. The non-zero solution of Eq.(\ref{int_eqn1}) determines the energy of trimer states.

However, in general, solving the coupled three-dimensional integral equation is highly nontrivial. Nevertheless, great simplification can be obtained in the subspace with ${\bf K_0}=0$. As shown in supplementary material \cite{supple}, for quantum state labelled by $(J,J_z)=(j+1/2,m+1/2)$ (where $j$ and $m$ are integers), $f_\sigma(\mathbf{k})$ satisfies
\begin{eqnarray}
f_\uparrow(\mathbf{k})&=&C^0_\uparrow f_0(k)Y_j^m(\Omega_\mathbf{k})+C^1_\uparrow f_1(k)Y_{j+1}^m(\Omega_\mathbf{k}),
 \nonumber \\
f_\downarrow(\mathbf{k})&=&C^0_\downarrow f_0(k)Y_j^{m+1}(\Omega_\mathbf{k})+C^1_\downarrow f_1(k)Y_{j+1}^{m+1}(\Omega_\mathbf{k}). \label{ansatz}
\end{eqnarray}
where $k=|{\bf k}|$ is the magnitude of ${\bf k}$ and $f_0, f_1$ are functions that only depend on $k$,
$C^0_\sigma, C^1_\sigma$ are Clebsch-Gordan coefficients,
\begin{eqnarray}
C^\delta_\sigma=\langle j+\delta,m-\sigma;\frac{1}{2},\sigma|j+\frac{1}{2},m+\frac{1}{2}\rangle,
\end{eqnarray}
with $\delta=0,1$ and $\sigma=\pm \frac{1}{2}$. After substituting Eq.(\ref{ansatz}) into
Eq.(\ref{int_eqn1}), Eq.(\ref{int_eqn1}) is reduced to two coupled one-dimension integral equations, whose explicit forms are given in supplementary material  \cite{supple} and can be solved numerically to determine trimer energy $E_3$.

\begin{figure}[t]
\includegraphics[width=2.7in]
{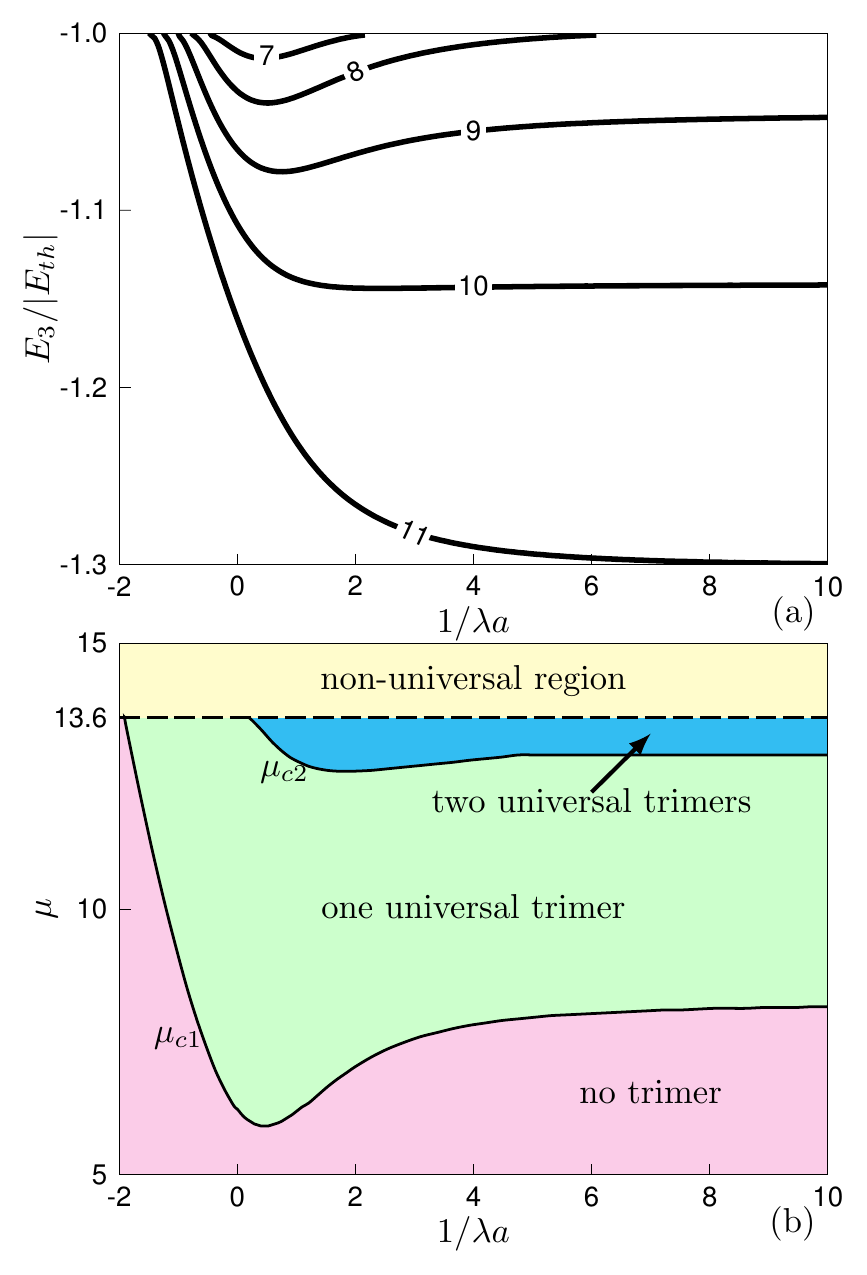}
\caption{(a) The ratio between $J=3/2$ trimer energy $E_3$ and atom-dimer threshold energy $|E_{th}|$, $\gamma=E_3/|E_{th}|$, as a function of $1/\lambda a$ for different mass ratios $M/m$ labelled in the curve. (b) The ``phase diagram" for $J=3/2$ trimer in term of $1/\lambda a$ and mass ratio $\mu=M/m$.  \label{ratio}}
\end{figure}

\textit{Results.} 
With SO coupling, the energies of $J=1/2$ channel and $J=3/2$ channel will split. Take $J=3/2$ channel as an example, the results are summarized as follows:

\textbf{1.} When $5.92\lesssim M/m<8.17$, there is no trimer state if there is no SO coupling. We find that with SO coupling, a trimer state will be induced in the vicinity of two-body resonance. It emerges from atom-dimer threshold at $a<0$ side and then merges into atom-dimer threshold at $a>0$ side, as shown in Fig. \ref{schematic}(a). The energy of such trimer state is independent of any high energy cutoff, thus, similar as universal ``Kartavtsev-Malykh" trimer, the ratio between trimer energy ($E_3$) and atom-dimer threshold energy ($E_\text{th}$) $\gamma=E_3/|E_{\text{th}}|$ is a universal function of $1/\lambda a$, as plotted in Fig 2(a). $\gamma<-1$ means that the trimer energy is below atom-dimer threshold.

\textbf{2.} When $8.17<M/m<13.6$, there exists at least one universal ``Kartavtsev-Malykh" trimers at positive $a$ side if there is no SO coupling. We find that with SO coupling, the lowest trimer starts to appear at $a<0$ side.
This trimer energy is also universal. The ratio $\gamma$ plotted in Fig 2(a) shows that $\gamma<-1$ from certain point with negative $a$ and saturates to a constant (the same value as predicted by Kartavtsev and Malykh for case without SO coupling) for large $1/\lambda a$. When $12.9\lesssim M/m<13.6$, an second trimer emerges at $a>0$ side.

\textbf{3.} When $M/m>13.6$, ``Thomas collapse" happens. Without SO coupling, there exists infinite number of Efimov trimers at resonance. SO coupling can not prevent ``Thomas collapse" \cite{Thomas}. While some shallow Efimov trimers disappears as SO coupling strength increases, their energies no longer obey universal scaling.

\begin{figure}[t]
\includegraphics[width=3.5in]
{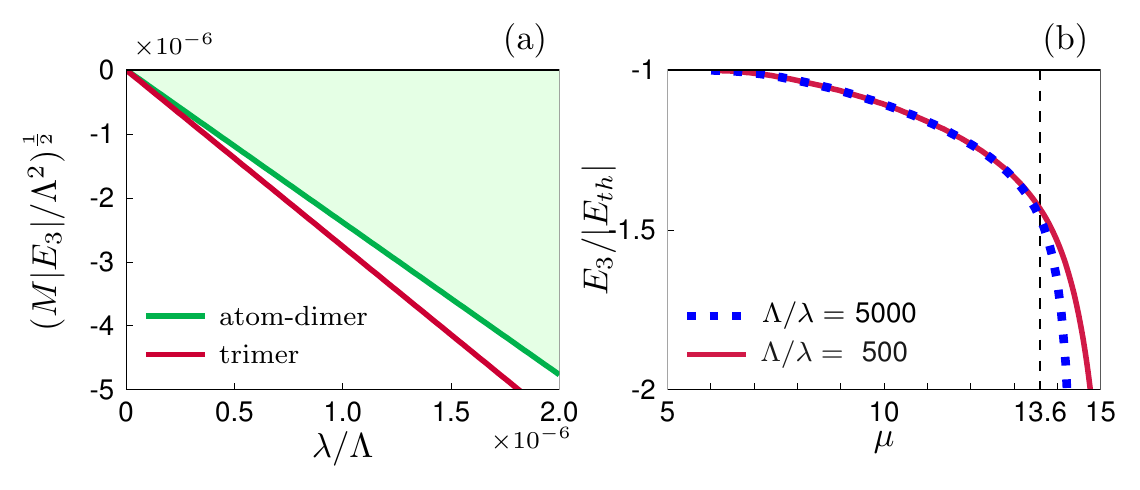}
\caption{(a) Trimer energy (in unit of high energy cutoff $\hbar^2/(M\Lambda^2)$) as a function of $\lambda/\Lambda$ for a given mass ratio $M/m=12$. (b) The lowest trimer energy $E_3$ (in unit of atom-dimer threshold energy $|E_{th}|$) as a function of mass ratio $\mu=M/m$ for two different high energy cutoff $\Lambda$. Both are plotted at two-body resonance $a=\infty$. \label{universal} }
\end{figure}

With results above, a ``phase diagram" for $J=3/2$ trimer is constructed in terms of dimensionless interaction parameter $1/\lambda a$ and mass ratio $\mu=M/m$, as shown in Fig. \ref{ratio}(b), where $\mu_\text{c1}$ ($\mu_\text{c2}$) is the critical mass ratio for the emergence of the first (second) universal trimer. It is interesting to note that $\mu_\text{c1}$ is a non-monotonic function of $1/\lambda a$, which reaches its minimum when $1/\lambda a$ is close to zero.

In Fig. \ref{universal}, we show that the trimer energy is indeed universal when $M/m<13.6$. At resonance, if the trimer energy is universal, $1/\lambda$ becomes the only length scale in the problem and trimer energy has to scale with $\hbar^2/(M\lambda^2)$. This scaling behavior is shown in Fig. \ref{universal}(a). In Fig. \ref{universal}(b), we plot the lowest trimer energy at resonance as a function of mass ratio, with two different high-energy cutoff $\Lambda$. It clearly shows that for $M/m<13.6$ the energy is independent of cutoff while it is not for $M/m>13.6$. The scenario of how universal ``Kartavtsev-Malykh" trimers crossover to Efimov trimer is similar to what has been discussed in Ref. \cite{Ueda} for the case without SO coupling.

\begin{figure}[t]
\includegraphics[width=2.5in]
{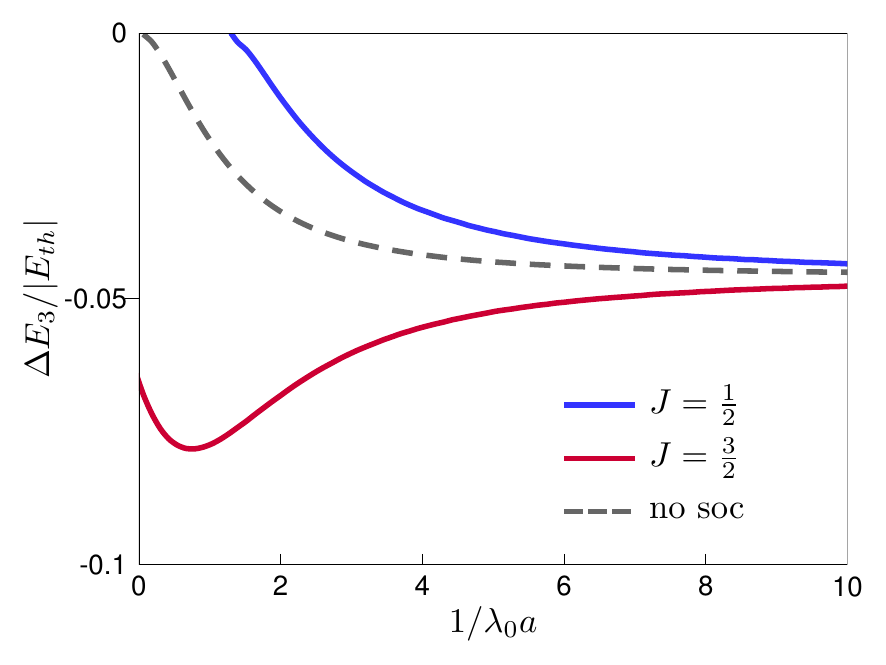}
\caption{The trimer binding energy $\Delta E_3 $ (in unit of atom-dimer threshold energy $|E_{\text{th}}|$ with a given $\lambda_0$) as a function of interaction strength $1/\lambda_0 a$ for a given mass ratio $M/m=12$. Dashed line represents the case without SO coupling $\lambda=0$ (six-fold degenerate). Two solid lines represent the cases for $J=3/2$ trimers (four-fold degenerate) and $J=1/2$ trimers (two-fold degenerate), respectively, for a given SO coupling strength $\lambda_0$.   \label{splitting}}
\end{figure}

Among above results ${\bf 1}-{\bf 3}$, ${\bf 1}$ and ${\bf 2}$ are most significant ones. It means that SO coupling favors trimer formation, i.e. universal trimer can now exist for a smaller mass ratio and also at $a<0$ side. Another way to view it is that, once $M/m\gtrsim 5.92$, trimer state can always be induced by increasing the strength of SO coupling, even for the system at weak interaction regime.

We attribute the reason that SO coupling favors trimer formation to the lifting of ground state degeneracy. If no SO coupling, all the bound states are highly degenerate, while SO coupling mixes different orbital angular momentum channels, which breaks such degeneracy and lowers the ground-state energy according to second perturbation theory. For example, in Fig. \ref{splitting}, we show a case with $M/m>8.17$ at $a>0$ side. The dashed line represents the energy of ``Kartavtsev-Malykh" trimer without SO coupling, where $J=3/2$ and $J=1/2$ states are degenerate. With SO coupling, it is found that the splitting between $J=3/2$ and $J=1/2$ increases the energy of $J=1/2$ trimer but lowers the energy of $J=3/2$ trimers. Consequently, the $J=3/2$ trimers can exist for a smaller mass ratio and also at $a<0$ side. Furthermore, because of the mixing of different orbital angular momentum channels is an intrinsic effect of SO coupling, we anticipate that our results qualitatively hold for a general type of SO coupling.

We also like to point out a common feature between three-body physics and two-body physics, that is, with SO coupling, physics at $a<0$ side becomes similar as conventional case at $a>0$ side. In two-body physics, for the type of SO coupling we considered here, dimer can form at $a<0$ side. While in three-body physics, universal `Kartavtsev-Malykh" trimer can now form at $a<0$ side. Both features are reminiscent of conventional case with $a>0$.

\textit{Final Remark.} Our results can potentially influence many-body physics. When the trimer energies touch the atom-dimer threshold, it will lead to an atom-dimer resonance where atom-dimer scattering length will change dramatically. Without SO coupling, usually the resonance position of Efimov trimer is controlled by three-body parameter $1/\Lambda a$, which is not tunable for a given mixture. While with SO coupling, the resonance position of universal trimer is controlled by $1/\lambda a$, which can be tuned quite flexibly  by SO coupling strength $\lambda$. Thus, this introduces a new way to manipulate a strongly interacting quantum many-body system.

{\it Acknowledgements.} We thank Zhenhua Yu and Ran Qi for helpful discussions. This work is supported by Tsinghua University Initiative Scientific Research Program, Chinese Academy of Sciences (XC), NSFC under Grant No. 11004118 (HZ) and No. 11174176 (HZ), No. 11104158 (XC) and No. 11374177 (XC), and NKBRSFC under Grant No. 2011CB921500.

\begin{widetext}

\vspace{0.05in}

\section{Supplementary Material}

In this supplementary material we present some details of solving two-body and three-body problem in the presence of spin-orbit coupling considered in the main text.

\subsection{I. Two-body calculation}

In the two-body part we consider an $\alpha$-atom (denoted as \textbf{1}) interacting with an $\beta$-atom (as \textbf{2}), and $\beta$ atom is with isotropic spin-orbit coupling. The Hamiltonian reads,
\begin{eqnarray}
H&=&H_0+U=\frac{\mathbf{p_1}^2}{2M}+\frac{(\mathbf{p_2}-\lambda\hat\sigma)^2}{2m}
+U,
 \\ \nonumber
 U&=&g\delta(\mathbf{r_1-r_2})\mathbf{I}. \qquad \sigma =\uparrow,\downarrow.
\end{eqnarray}
A two-body state with total momentum $K$ is assumed as
 \begin{eqnarray}
|\Psi\rangle=\sum_{{\bf p},\sigma} \Psi_\sigma(\mathbf{K-p,p})  |\mathbf{K-p,p};\sigma\rangle,
\end{eqnarray}
and we solve the problem through Lippman-Schwinger equation
\begin{eqnarray}
|\Psi\rangle=\frac{1}{E-H_0}U|\Psi\rangle.
\end{eqnarray}
Multiply $\langle\mathbf{K-p,p};\sigma|$ on each side, we get
\begin{eqnarray}
\Psi_\sigma(\mathbf{K-p,p})=
g \sum_{\mathbf{q},\sigma'}G_{\sigma\sigma'}(\mathbf{K-p,p}) \Psi_{\sigma'}(\mathbf{K-q,q}), \label{two_body_eqn}
\end{eqnarray}
where $G_{\sigma\sigma'}$ is the two-particle Green's function,
\begin{eqnarray}
G_{\uparrow\uparrow}(\mathbf{k_1,k_2})&=&\frac{\cos^2\frac{\theta_\mathbf{k_2}}{2}}
{E-\epsilon_{\mathbf{k_1}}-\epsilon^+_{\mathbf{k_2}}}+
\frac{\sin^2\frac{\theta_\mathbf{k_2}}{2}}
{E-\epsilon_{\mathbf{k_1}}-\epsilon^-_{\mathbf{k_2}}},\\ \nonumber
G_{\downarrow\downarrow}(\mathbf{k_1,k_2})&=&\frac{\sin^2\frac{\theta_\mathbf{k_2}}{2}}
{E-\epsilon_{\mathbf{k_1}}-\epsilon^+_{\mathbf{k_2}}}+
\frac{\cos^2\frac{\theta_\mathbf{k_2}}{2}}
{E-\epsilon_{\mathbf{k_1}}-\epsilon^-_{\mathbf{k_2}}},\\ \nonumber
G_{\downarrow\uparrow}=G^*_{\uparrow\downarrow}&=&\sin\frac{\theta_\mathbf{k_2}}{2}\cos\frac{\theta_\mathbf{k_2}}{2}
e^{i\phi_\mathbf{k_2}}(\frac{1}{E-\epsilon_{\mathbf{k_1}}-\epsilon^+_{\mathbf{k_2}}}-
\frac{1}{E-\epsilon_{\mathbf{k_1}}-\epsilon^-_{\mathbf{k_2}}}),
\end{eqnarray}
with $\epsilon_\mathbf{k}=k^2/2M$ and $\epsilon^\pm_\mathbf{p}=(|p|\pm\lambda)^2/2m$. Here $\theta_{\mathbf{k}}$ and $\phi_{\mathbf{k}}$ stand for the polar angle and azimuth angle of $\mathbf{k}$.

To solve this equation, we define $f_{\sigma}$ as
\begin{eqnarray}
f_\sigma=g\sum_{\mathbf{p}}\Psi_\sigma(\mathbf{K-p,p}).
\end{eqnarray}
With such definition we simplify equation (\ref{two_body_eqn}) into
\begin{eqnarray}
\Psi_\sigma(\mathbf{K-p,p})=\sum_\sigma G_{\sigma\sigma'}(\mathbf{p,K-p})
f_{\sigma'}.
\end{eqnarray}
After a summation over $\mathbf{p}$, we get a close equation of $f_\sigma$,
\begin{eqnarray}
f_\sigma=g\sum_{\sigma',\mathbf{p}}G_{\sigma\sigma'}(\mathbf{p,K-p})
f_{\sigma'}.
\end{eqnarray}
The above equation can be written down in a matrix form,
\begin{eqnarray}
\left(
  \begin{array}{cc}
    \sum_{\mathbf{p}}G_{\uparrow\uparrow}-1/g & \sum_{\mathbf{p}}G_{\uparrow\downarrow} \\
    \sum_{\mathbf{p}}G_{\downarrow\uparrow} &  \sum_{\mathbf{p}}G_{\downarrow\downarrow}-1/g\\
  \end{array}
\right)
\left(
  \begin{array}{c}
    f_\uparrow \\
    f_\downarrow \\
  \end{array}
\right)=0.
\end{eqnarray}
We can simplify the summation in above equation into
\begin{eqnarray}
\left(
  \begin{array}{cc}
    D+Y\cos\theta & Y\sin\theta e^{-i\phi} \\
    Y\sin\theta e^{i\phi} &  D-Y\cos\theta\\
  \end{array}
\right)
\left(
  \begin{array}{c}
    f_\uparrow \\
    f_\downarrow \\
  \end{array}
\right)=0.
\end{eqnarray}
Here $\theta$ and $\phi$ stand for $\theta_{\mathbf{K}}$ and $\phi_{\mathbf{K}}$,
$D$ and $Y$ are defined as following ($K=|\mathbf{K}|$, $\mu=M/m$),
\begin{eqnarray}
D&=&-\frac{1}{2\pi(1+\mu)}[\frac{1}{a}-
\frac{(K+\mu\lambda)\sqrt{\mu(K-\lambda)^2-2(1+\mu)ME}+
(K-\mu\lambda)\sqrt{\mu(K+\lambda)^2-2(1+\mu)ME}}{2K(1+\mu)}],\\ \nonumber
Y&=&-\frac{1}{12\pi(1+\mu)^2K^2}
\{[\mu(K^2+K\lambda+\lambda^2)+3K^2-2(1+\mu)ME]\sqrt{\mu(K-\lambda)^2-2(1+\mu)ME}\\ \nonumber
&&-[\mu(K^2-K\lambda+\lambda^2)+3K^2-2(1+\mu)ME]\sqrt{\mu(K+\lambda)^2-2(1+\mu)ME}
\}.
\end{eqnarray}
Finally, the two-body binding energy is determined by following algebraic equation,
\begin{eqnarray}
D^2-Y^2=0.
\end{eqnarray}

\subsection{II. Three-body calculation}

We use the same technique to solve the three-body problem, the Lippman-Schwinger equation in
momentum space is given by,
\begin{eqnarray}
\Psi_\sigma(\mathbf{k_1,k_2,K_0-k_1-k_2})&=& g
\sum_{\mathbf{q},\sigma'}G_{\sigma\sigma'}(\mathbf{k_1,k_2,K_0-k_1-k_2}) \nonumber \\
&\times&[\Psi_{\sigma'}(\mathbf{q,k_2,K_0-q-k_2})+\Psi_{\sigma'}(\mathbf{k_1,q,K_0-q-k_1})],
\label{eqn_Psi}
\end{eqnarray}
with $\mathbf{K_0}$ the total momentum and $G_{\sigma\sigma'}$ the Green's function of three free particles.
Like we have done in two-body part, defining an auxiliary function
$f_\sigma(\mathbf{p})=g\sum_\mathbf{q}\Psi_\sigma(\mathbf{q,K_0-p,p-q})=-g\sum_\mathbf{q}\Psi_\sigma(\mathbf{K_0-p,q,p-q})$, we simplify equation (\ref{eqn_Psi}) into
\begin{eqnarray}
f_\sigma(\mathbf{k})=g\sum_{\mathbf{p},\sigma'}G_{\sigma\sigma'}(\mathbf{p,K_0-k,k-p})
[f_{\sigma'}(\mathbf{k})- f_{\sigma'}(\mathbf{K_0-p})]. \label{eqn_f}
\end{eqnarray}

Since the total angular momentum operator $\mathbf{J}$ can be represented as
\begin{eqnarray}
\mathbf{J}&=&\mathbf{r_1\times p_1+r_2\times p_2 +r_3\times p_3}\\ \nonumber
&=& i\mathbf{\partial_{p_1}\times p_1}+i\mathbf{\partial_{p_2}\times p_2}+i\mathbf{\partial_{p_3}\times p_3}.
\end{eqnarray}
If we restrict the calculation in total translational momentum $\mathbf{K_0}=0$ and total angular momentum $(J,m_J)=(j+1/2,m+1/2)$
sub-Hilbert space.
It is easy to prove that the wave function in momentum space takes the form,
\begin{eqnarray}
\Psi_\uparrow(\mathbf{k_1,k_2,k_3})&=&\delta(\mathbf{k_1+k_2+k_3})\sum_{j_1,j_2,J'}\varphi_{j_1,j_2,J'}(k_1,k_2)\times \\ \nonumber
&&\sum_{m_1,m_2,m_J'}
\langle j_1,m_1;\frac{1}{2},\frac{1}{2}|J',m_J'\rangle\langle j_2,m_2;J',m_J'|j+\frac{1}{2},m+\frac{1}{2}\rangle Y_{j_1}^{m_1}(\Omega_\mathbf{k_1})
Y_{j_2}^{m_2}(\Omega_\mathbf{k_2}), \\ \nonumber
\Psi_\downarrow(\mathbf{k_1,k_2,k_3})&=&\delta(\mathbf{k_1+k_2+k_3})\sum_{j_1,j_2,J'}\varphi_{j_1,j_2,J'}(k_1,k_2)\times \\ \nonumber
&&\sum_{m_1,m_2,m_J'}
\langle j_1,m_1;\frac{1}{2},-\frac{1}{2}|J',m_J'\rangle\langle j_2,m_2;J',m_J'|j+\frac{1}{2},m+\frac{1}{2}\rangle Y_{j_1}^{m_1}(\Omega_\mathbf{k_1})
Y_{j_2}^{m_2}(\Omega_\mathbf{k_2}), \\ \nonumber
\end{eqnarray}
where $k_i$ is the magnitude of $\mathbf{k}_i$, $\varphi_{j_1,j_2,J'}$ is an undetermined function depends on $k_1$ and $k_2$, and $Y_j^m(\Omega_\mathbf{k})$ stands for spherical harmonics $Y_j^m(\theta_{\mathbf{k}},\phi_{\mathbf{k}})$.

Therefore, after summing over $\mathbf{k_1}$, only $j_1=m_1=0$ terms contribute to $f_\sigma$, and we obtain
our ansatz for $f_\sigma$:
\begin{eqnarray}
f_\uparrow(\mathbf{k})&=&\langle j,m;\frac{1}{2},\frac{1}{2}|j+\frac{1}{2},m+\frac{1}{2}\rangle f_0(k)Y_j^m
+\langle j+1,m;\frac{1}{2},\frac{1}{2}|j+\frac{1}{2},m+\frac{1}{2}\rangle f_1(k)Y_{j+1}^{m} \\ \nonumber
f_\downarrow(\mathbf{k})&=&\langle j,m+1;\frac{1}{2},-\frac{1}{2}|j+\frac{1}{2},m+\frac{1}{2}\rangle f_0(k)Y_j^{m+1}
+\langle j+1,m+1;\frac{1}{2},-\frac{1}{2}|j+\frac{1}{2},m+\frac{1}{2}\rangle f_1(k)Y_{j+1}^{m+1},\\ \nonumber
\end{eqnarray}
where $f_0$ and $f_1$ are two functions only depend on the magnitude of $\mathbf{k}$ and $Y_j^m$ is short for $Y_j^m(\Omega_\mathbf{k})$.

After substituting the ansatz into equation (\ref{eqn_f}), we find following equation,
\begin{eqnarray}
Z(k)
\left(
\begin{array}{c}
      f_0(k) \\
      f_1(k)
    \end{array}
\right)
=\int_0^\infty dp K_j(k,p)
\left(
\begin{array}{c}
  f_0(p) \\
  f_1(p)
\end{array}
\right). \label{int_eqn}
\end{eqnarray}
The matrix elements of $Z$ and $K_j$ are given by,
\begin{eqnarray}
K_j&=&\frac{p^2}{(2\pi)^3}\left(
                            \begin{array}{cc}
                              W_j & -kR_{j+1}-pR_j \\
                              -kR_j-pR_{j+1} & W_{j+1} \\
                            \end{array}
                          \right),\\ \nonumber
Z_{11}=Z_{22}&=&-\frac{1}{2\pi(1+\mu)}[\frac{1}{a}\\ \nonumber&&-\frac{(k+\mu\lambda)\sqrt{\mu(k-\lambda)^2+(1+\mu)(k^2-2ME)}+
(k-\mu\lambda)\sqrt{\mu(k+\lambda)^2+(1+\mu)(k^2-2ME)}}{2(1+\mu)k}], \\ \nonumber
Z_{12}=Z_{21}&=&\frac{1}{12\pi(1+\mu)^2k^2}
\{[\mu(2k^2+k\lambda+\lambda^2)+4k^2-2(1+\mu)ME]\sqrt{\mu(k-\lambda)^2+(1+\mu)(k^2-2ME)}\\ \nonumber
&&-[\mu(2k^2-k\lambda+\lambda^2)+4k^2-2(1+\mu)ME]\sqrt{\mu(k+\lambda)^2+(1+\mu)(k^2-2ME)}
\},\\ \nonumber
\end{eqnarray}
with $W_j$ and $R_j$ defined as
\begin{eqnarray}
W_j(k,p)&=&\frac{4\pi}{2j+1}\int\sin\theta d\theta P_j(\cos\theta)\times[\frac{1}{2E-k^2-p^2-\mu(|\mathbf{k+p}|+\lambda)^2}
+\frac{1}{2E-k^2-p^2-\mu(|\mathbf{k+p}|-\lambda)^2}],\\ \nonumber
R_j(k,p)&=&\frac{4\pi}{2j+1} \int \frac{\sin\theta d\theta P_j(\cos\theta)}{|\mathbf{k+p}|}\times
[\frac{1}{2E-k^2-p^2-\mu(|\mathbf{k+p}|+\lambda)^2}
-\frac{1}{2E-k^2-p^2-\mu(|\mathbf{k+p}|-\lambda)^2}],
\end{eqnarray}
where $P_j$ is the j-th Legendre polynomial and $\theta$ represents the angle between vectors $\mathbf{k}$ and $\mathbf{p}$. Both $K_j$ and $Z$ are dependent on $E$. Only when $E$ is equal to a three-body bound state energy, there exists nonzero solution of equation (\ref{int_eqn}). This helps us to determine the energies of trimers.

\end{widetext}


\begin{thebibliography}{99}
\bibitem{magnetic}
If SO coupling is generated by Raman scheme as used in current experiment, the heating due to spontaneous mission is significant for light atom like ${}^6$Li because of small hyperfine splitting of excited states \cite{MIT}. While recent works Ref. \cite{spielman_3d_2,Xu} propose to generate SO coupling using magnetic field pulse without involving light, which can avoid heating problem for the light element.
\bibitem{Thomas}
Since the kind of SO coupling we considered is linear in $k$, it will not change the short-range behavior dominated by $k^2$ kinetic term. Therefore, the short-range behavior of the three-body potential, which scales as $-1/R^2$ ($R$ is the hyper-radius of three atoms) and leads to Efimov physics, will not be affected.
\end{thebibliography}
\end{document}